\documentclass[aps,pra,twocolumn,a4paper,amsmath,amssymb,showpacs,%
floatfix]{revtex4}
\usepackage{graphicx,epstopdf}
\usepackage{braket}

\newcommand{\Fkt}[1]{\,\mathsf {#1}}
\ifx\Tr\renewcommand{\Tr}{\Fkt{Tr}}
\else\newcommand{\Tr}{\Fkt{Tr}}
\fi

\begin{document}

\title{Minimum number of input states required for quantum gate
  characterization} 

\author{Daniel M. Reich}
\affiliation{Theoretische Physik, Universit\"{a}t Kassel,
  Heinrich-Plett-Str. 40, D-34132 Kassel, Germany} 
\author{Giulia Gualdi} 
\altaffiliation[present address: ]{QSTAR, Largo Enrico Fermi 2, I-50125 Florence, Italy}
\affiliation{Theoretische Physik, Universit\"{a}t Kassel,
  Heinrich-Plett-Str. 40, D-34132 Kassel, Germany}
\author{Christiane P. Koch}
\affiliation{Theoretische Physik, Universit\"{a}t Kassel,
  Heinrich-Plett-Str. 40, D-34132 Kassel, Germany} 
\email{christiane.koch@uni-kassel.de}
\date{\today}

\begin{abstract}
  We derive an algebraic framework which identifies the minimal
  information required to assess how well a quantum device
  implements a desired quantum operation. Our approach is based on
  characterizing only the unitary part of an open system's
  evolution. We show that a reduced set of input states is sufficient to
  estimate the average fidelity of a quantum gate, avoiding a
  sampling over the full Liouville space. Surprisingly, the minimal
  set consists of only two input states, independent of the Hilbert space
  dimension. The minimal set is, however, impractical for device
  characterization since one of the
  states is a totally mixed thermal state and extracting bounds for
  the average fidelity is impossible. We therefore present two further
  reduced sets of input states that allow, respectively, for numerical
  and analytical bounds on the average fidelity.
\end{abstract}

\maketitle

\section{Introduction}
\label{sec:intro}

The usual measure to  assess how well a quantum device
implements a desired quantum operation is the average fidelity, 
\begin{eqnarray}
  \label{eq:Fav}
  F_{av}  = \int\Braket{\Psi|O^{+}\mathcal{D}
    \left(\Ket{\Psi}\Bra{\Psi}\right)O|\Psi}d\Psi \,,
\end{eqnarray}
where $O$ denotes the desired unitary and the actual time evolution is
described by the dynamical map $\mathcal D$.
The standard approach to determine $F_{av}$ relies on quantum process
tomography~\cite{NielsenChuang}. 
In practice, the  average fidelity of a quantum process in a
$d$-dimensional Hilbert space is often estimated by performing   
quantum state tomography in a $d^2$-dimensional Hilbert
space. For $N$ qubits $d=2^{N}$. The fidelity can also be obtained 
by determining the  process matrix which is of
size $d^2\times d^2$.  In both cases
quantum process tomography scales exponentially in 
resources~\cite{MohseniPRA08}. 
For quantum devices to be realized and tested in practical
applications, a less resource-intensive approach to characterization
is required. 

Recent attempts at reducing the required resources employ
stochastic sampling of the input states and measurement
observables~\cite{FlammiaPRL11,daSilvaPRL11,ShabaniPRL11,SchmiegelowPRL11}.
The process matrix can be estimated efficiently if it is
sparse in a known
basis~\cite{MohseniPRA09,ShabaniPRL11,SchmiegelowPRL11}. For general  
unitary operations and without assuming any prior knowledge, 
Monte Carlo sampling to determine state fidelities in the
$d^2$-dimensional Hilbert space currently seems to be the most
efficient approach~\cite{FlammiaPRL11,daSilvaPRL11,SteffenPRL12}. 
This is due to the fact that the approach directly targets 
the fidelity between the desired operation and the implemented process 
rather than fully characterizing the process and subsequently
comparing it to the desired operation. It also comes
with the advantage of separable input states and local measurements.
For $N$ qubits, this approach requires the ability to prepare 
$6^N$ input states since there are 6 eigenstates for the three Pauli
operators for each qubit and the ability to measure all of the 
$d^2=2^{2N}$ operators that form an orthonormal Hermitian operator
basis.  

Another approach to the estimation of the average fidelity 
exploits its property of being a second-degree polynomial in the
states, utilizing a so-called two-design~\cite{BenderskyPRL08}. 
A commonly used two-design is made up of the $d(d+1)$ states of $d+1$ 
mutually unbiased bases.
The average fidelity is then written as a sum over state fidelities
for these states~\cite{BenderskyPRL08}. The latter implies preparation
of entangled input states since only three out of the $d+1$ mutually
unbiased bases consist of separable states~\cite{LawrencePRA11}. 
Both Monte Carlo characterization and the two-design approach yield
the average fidelity with an arbitrary,  
prespecified accuracy. 
Alternatively, bounds on the average fidelity can be obtained from two
classical fidelities~\cite{HofmannPRL05} where each classical fidelity is
expressed as a sum over $d$ state fidelities. 
The different requirements of Monte Carlo characterization, the
two-design approach and the two classical fidelities in terms of the
number of input states raise the question of what is the minimal set
of states to determine $F_{av}$.

Here we show that a minimal set of states can be identified by the
requirements to allow for distinguishing any two unitaries and assess
whether the time evolution is unitary. We find the minimal set of
states to consist of only two states, independent of the size of
Hilbert space. The minimal set contains, however, a totally mixed
thermal state which is impractical for  
experiments. We therefore also introduce a reduced sets of states that
consist of the minimum number of pure states required to distinguish any
two unitaries and assess whether the time evolution is unitary.
The average fidelity can then be  estimated by evaluating a distance
measure for the reduced set of states. The corresponding 
protocol consists of preparing 
$d+1$ pure states, defined in $d$-dimensional Hilbert space,
and measuring the corresponding state fidelities. 
We show numerically that the estimate of the gate
error differs from $F_{av}$ by less than a factor 2.5 in the worst
case and  on average by a factor 1.2.  
We furthermore demonstrate that evaluation of state
fidelities for the reduced set of states is also sufficient to
quantify the non-unitarity of the process. 
This allows to determine 
whether the gate error is due to decoherence or due to unitary errors
that are easier to mitigate. 

If analytical instead of numerical bounds on the average fidelity are
desired, the reduced set needs to contain $2d$ states, i.e., 
our approach generalizes the estimate of the average fidelity in terms
of two classical fidelities~\cite{HofmannPRL05}. We show that the
specific states utilized by Refs.~\cite{HofmannPRL05,LanyonSci11}
also fulfill the requirements for distinguishing any two unitaries and
assessing unitarity of the time evolution -- as do the states of any two
mutually unbiased bases. 

Our paper is organized as follows: The algebraic framework
for identifying the minimum requirements to distinguish any two
unitaries and assess unitarity of the time evolution is derived in
Section~\ref{sec:algframe}, introducing the concepts of commutant
space and total rotation. 
Section~\ref{sec:redset} presents the
reduced sets of states and discusses their use for extracting an
estimate of the average fidelity. 
The relationship of our approach to the two classical fidelities of
Ref.~\cite{HofmannPRL05} is established in Section~\ref{sec:Hofmann},
and our results are summarized in Section~\ref{sec:concl}. 
Detailed proofs of the claims made in Section~\ref{sec:algframe} are
provided in Appendix~\ref{sec:app}.

\section{Algebraic framework:
  Commutant space and total rotation}
\label{sec:algframe}
To identify the reduced set of states, 
we introduce the concepts of  commutant space of a set of density 
operators and total rotation. 
We assume purely coherent time evolution with an unknown
unitary $U \in U(d)$, such that $\rho(T)=\mathcal D(\rho)=U\rho U^+$,
and generalize later to non-unitary time evolution.
Since the evolution is insensitive to a
global phase, $U$ is an element of  
the projective unitary group, $PU(d)$, i.e., the quotient $U(d)/U(1)$
of the unitary groups $U(d)$ and $U(1)$.  
Given a set of states, $\{\rho_j=\rho_j(t=0)\}$, we consider the 
map $\mathcal M: PU(d) \longrightarrow \bigoplus_{j}\mathbb{C}^{d\times
  d}$, mapping the unitary $U$ onto the set of time-evolved states,
$\{\rho^U_j(T)=U\rho_jU^+\}$. We can differentiate any two unitaries
$U$, $U'$ if and only if the map $\mathcal M$ is injective. We show
that $\mathcal{M}$ is injective if the commutant space of
the set 
$\{\rho_j\}$ has only one element,
the identity. 

We define
the commutant space $K(\rho)$ of a single density operator $\rho$ 
as the set of all linear operators in $PU(d)$ that commute with $\rho$. It
contains the identity and all operators that have a common eigenbasis
with $\rho$. Unitaries $\tilde U$ 
in the commutant space of $\rho$ cannot 
be distinguished from $\openone$ by time-evolving $\rho$
since $\tilde U\rho \tilde U^+=\tilde U \tilde
U^+\rho=\rho$. Therefore, to distinguish a unitary $U$ from the
identity, the time evolution of at least two density operators with
different eigenbases is required. 
Once we can differentiate
an arbitrary unitary from the identity, we can differentiate it from
any other unitary (and $\mathcal M$ is injective). This follows from
the fact that $PU(d)$ is a group. 
We define the commutant space of a set of density 
operators, $\mathcal{K}(\{\rho_j\})$, as the intersection
of all  $K(\rho_j)$, i.e., the set of all linear operators that
commute with \textit{each} $\rho_j$. Suppose the identity is the only
element of the commutant space $\mathcal{K}(\{\rho_j\})$. 
Then the identity is the only time evolution that leaves \textit{all}
$\rho_j$ unchanged and 
we can distinguish the identity from 
all other time evolutions by inspecting the time-evolved states. 
The detailed proof that injectivity of $\mathcal M$ is
equivalent to  $\mathcal{K}(\{\rho_j\})$ having identity as its only
element is given in Appendix~\ref{sec:app}.

In order to determine the states of the reduced set 
$\{\rho_j\}$ that have a commutant space  $\mathcal{K}(\{\rho_j\})$
with identity as its only element, we introduce the concept of total
rotation. Unitary evolution corresponds to rotations in Hilbert
space. Spanning the Hilbert space by an arbitrary complete orthornomal
basis $\{|\varphi_i\rangle\}$, a complete set of $d$
one-dimensional orthonormal projectors is obtained, $\mathcal
P_c\equiv\{P_i=|\varphi_i\rangle\langle\varphi_i|\}$. We construct 
density operators within this basis, for example by choosing a single
state, $\rho_B=\sum_{i=1}^d\lambda_iP_i$ with $\lambda_i\neq
\lambda_j$ for $i\neq j$, or a set of $d$ states,
$\{\rho_{B,i}\}$, $\rho_{B,i}=P_i$, $i=1,\ldots, d$. The time-evolved
basis state $\rho_B(T)$ or states 
$\{\rho_{B,i}(T)\}$ allow for distinguishing all those unitaries from
identity that do not have common eigenspaces with all $P_i$. To
distinguish the remaining unitaries from identity, we construct one
additional state, $\rho_{TR}$, that is guaranteed to have no
common eigenspace with any $P_i$. This is achieved by introducing a
totally rotated one-dimensional projector $P_{TR}$ obeying
$P_{TR}P_i\neq 0$ $\forall P_i \in \mathcal P_c$ and taking
$\rho_{TR}=P_{TR}$. Adding $P_{TR}$ to $\mathcal P_c$ makes the set of
projectors complete and totally rotating, 
$\mathcal P_{cTR}=\mathcal P_c \cup\{P_{TR}\}$.
A set of states $\{\rho_j\}$
is complete and totally rotating if the subset of the projectors onto
the one-dimensional eigenspaces of the $\{\rho_j\}$ is complete and
totally rotating. For example, $\{\rho_B,\rho_{TR}\}$ or
$\{\rho_{B,1},\ldots,\rho_{B,d},\rho_{TR}\}$. 
We show in Appendix~\ref{sec:app} that the identity
is the only projective unitary operator that has a common eigenbasis
with all elements of such a set of states.

We have thus constructed a reduced set of states $\{\rho_j\}$
that allows for differentiating any two unitaries by inspection of the
time-evolved states, $\{\rho_j(T)\}$. 
For coherent time evolution, we can evaluate 
\begin{eqnarray}
  \label{eq:Fredi}
  F_{j} = \Tr\left[\rho^O_{j}\rho_{j}(T)\right] \,,
\end{eqnarray}
which matches each state $\rho_j$, subjected to the ideal operation,
$\rho_j^O=O\rho_{j}O^+$, to the actually evolved state,
$\rho_j(T)=\mathcal{D}(\rho_{j})$, for
all $\rho_j(T)$. A suitable combination of the resulting $F_j$
yields an estimate of $F_{av}$.  
However, for a possibly incoherent time evolution,
we need to quantify the 'non-unitarity' of the actual evolutions
$\mathcal D(\rho_j)$. This can be done by checking whether $\mathcal D$
maps projectors onto projectors, reflecting rotations
in Hilbert space. 
We show in Appendix~\ref{sec:app}
that indeed unitarity of a dynamical
map $\mathcal D$ is equivalent to $\mathcal D$ mapping (i) a set
$\{P_i\}$ of $d$ one-dimensional orthogonal projectors onto another
such set 
$\{\tilde P_i\}$ of $d$ one-dimensional orthogonal projectors and (ii)
a projector $P_{TR}$ that is totally rotated with respect to
the set $\{P_i\}$ onto a one-dimensional projector. 

\section{Reduced set of states yielding numerical bounds on the 
  average fidelity} 
\label{sec:redset}
A set of density operators that allows for
both differentiating any two unitaries and measuring the non-unitarity
of any dynamical map $\mathcal D$ is thus given by 
\begin{subequations}
  \label{eq:rho}
\begin{eqnarray}\label{eq:rho_Bi}
  \rho_{B,i} &=& |\varphi_i\rangle\langle\varphi_i| \,,
  \quad\quad\quad\quad\quad i=1,\ldots, d\,,\\
  \rho_{TR} &=& \frac{1}{d}\sum_{i,j=1}^d 
  |\varphi_i\rangle\langle\varphi_j|\,.
\end{eqnarray}
\end{subequations}
By construction, the states $\rho_{B,i}$, $\rho_{TR}$ 
are pure. 
They are separable if a separable basis is chosen, i.e., if
all $|\varphi_i\rangle$  are separable. 
Another suitable reduced set to differentiate any two unitaries and
measure  non-unitarity of $\mathcal D$ is given by 
\begin{eqnarray}
  \label{eq:minset}
  \{\rho_B=\sum_i\lambda_iP_i,\rho_{TR}\}\quad \mathrm{with}\quad
  \lambda_i\neq\lambda_j\quad \mathrm{for}\quad i\neq j\,.
\end{eqnarray} 
This is the minimal set~\footnote{The corresponding
  extension of the 
  proof requires $\mathcal D$ to be unital. Any distance measure based
  on $\{\rho_B=\sum_i\lambda_iP_i,\rho_{TR}\}$ must therefore contain an
  additional check whether $\mathcal D$ maps the identity onto itself.
  which can be performed by adding a suitable third state to the
  set.}. However, for the 
characterization of quantum gates, it is preferable to use the pure input
states defined in Eq.~\eqref{eq:rho}. Each of these states, when
evolved in time, is characterized, to leading order, by $d^2$ real
parameters. Knowledge of the total $d^2(d+1)$ parameters is 
sufficient to determine whether the time evolution matches 
the desired unitary. 

Note that both reduced sets are also sufficient to 
reconstruct a unitary that is close to a given open system
evolution. This implies that, in optimal control
calculations for quantum gates in the presence of decoherence,
propagation of two states, 
$\{\rho_B=\sum_i\lambda_iP_i,\rho_{TR}\}$ independent of the system size $d$, 
is sufficient. This reduces significantly 
the numerical effort compared to the 
$d^2$ states used to date~\cite{KallushPRA06}.

\subsection{Estimating the gate error}
\label{subsec:gateerror}
The usual figure of merit in quantum process
tomography, the  average fidelity,
$F_{av}$, or, respectively, the 
gate error, $1-F_{av}$, can be estimated by averaging over the 
distance measures $F_j$, Eq.~\eqref{eq:Fredi}, 
for each state $\rho_j$ in the reduced set.
Each $F_j$ becomes maximal if and only if $O\rho_{j}O^+=\mathcal
D\rho_{j}$. Our protocol thus consists in the preparation of $d+1$
states $\rho_j$, Eq.~\eqref{eq:rho}, and measurement of the
corresponding state fidelities, $F_j$, for the time-evolved states,
$\mathcal D(\rho_j)$. A possible choice of states is e.g.
\begin{equation}
  \label{eq:rho_OBi}
  \left(\rho^O_{B,i}\right)_{nm}\equiv\left(P_i\right)_{nm} 
  = \delta_{ni}\delta_{mi} 
\end{equation}
in the computational basis.  
The average over the $F_j$ can employ 
the arithmetic mean or a modified geometric mean, 
\begin{eqnarray}
  \label{eq:Farith}
  F^{arith}_{unitary} &=& \frac{1}{d+1}\left[ 
    \sum_{i=1}^d F_{B,i} + F_{TR}
  \right]\,,\\
  \label{eq:Fgeom}
  F^{geom}_{unitary}&=&\frac{1}{d+1}+
  \left(1-\frac{1}{d+1}\right) \left[ \prod_{i=1}^d F_{B,i}\;\cdot\;
    F_{TR}
  \right]\,,
\end{eqnarray}
or a combination of the two. The first term in Eq.~\eqref{eq:Fgeom}
ensures $F^{geom}_{unitary}$ to take values in 
the same interval, $[\frac{1}{d+1},1]$, as $F_{av}$ for unitary
evolution.
$F^{arith/geom}_{unitary}=1$ only for a purely coherent time
evolution that perfectly implements the desired gate $O$ for all
states in the reduced set. 
While the arithmetic mean weights all state fidelities linearly, the 
geometric mean works best if the error is due to a single $F_j$.  
An optimized way to extract information from all the
$F_j$ is obtained by a suitable combination of the arithmetic and
geometric mean: We
define a fidelity that switches from the arithmetic mean to the
geometric one, should the state fidelities for all the $\rho_{B,i}$ be
close to one,
\begin{subequations}  \label{eq:Fred}
\begin{eqnarray}
  F^{\lambda}_{unitary}=\lambda F^{geom}_{unitary} +
  \left(1-\lambda\right) F^{arith}_{unitary}
\end{eqnarray}
with 
\begin{eqnarray}\label{eq:lambda}
  \lambda = 1 - \frac{1-\prod_{i=1}^d F_{B,i}}
  {1-\prod_{i=1}^d F_{B,i}\;\cdot\; F_{TR}}\,.
\end{eqnarray}
\end{subequations}
The choice of $\lambda$ is motivated as follows: $\lambda=1$
such that $F^{\lambda}_{unitary}= F^{geom}_{unitary} $
if $F_{B,i}=1$ for all $i$, i.e., in cases where the gate error is
captured by $F_{TR}$ alone; and $\lambda=0$ yielding
$F^{\lambda}_{unitary}=F^{arith}_{unitary}$ if
$F_{TR}=1$, i.e., when the gate error is comprised in the $F_{B,i}$.  

\begin{figure}[t]
  \centering
  \includegraphics[width=0.95\linewidth]{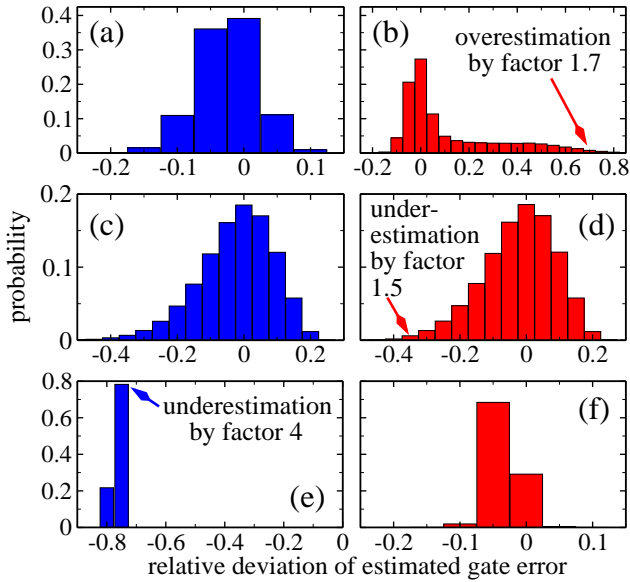}
  \caption{(Color online) 
    Probability of the estimated gate error's
    relative deviation from the standard gate error,
    $\Delta =
    (\varepsilon_{estim} - \varepsilon_{av})/\varepsilon_{av}$, 
    for 100.000
    realizations when using $F^{arith}_{unitary}$, 
    Eq.~\eqref{eq:Farith}, (left column) and  $F^{\lambda}_{unitary}$,
    Eq.~\eqref{eq:Fred}, (right column). Shown are the results  for
    randomized dynamical maps with $O=$CNOT 
    (a, b), truly random unitaries with $O=$CNOT (c-d) and randomized
    unitaries with $O=\openone$ (e-f). 
    Positive and negative values of $\Delta$, 
    corresponding to under- and overestimation of the gate error, do not
    scale equivalently. 
    The scale for overestimation ($\Delta>0$) ranges from zero to
    infinity while that for underestimation ($\Delta<0$) 
    is confined to $[-1,0)$.
  }
  \label{fig:histo}
\end{figure}
Figure~\ref{fig:histo}(a)
shows the probability of obtaining a certain relative deviation of
the estimated gate error for randomized dynamical maps
and CNOT as the target gate. 
The randomized dynamical maps were obtained by creating a random
matrix~\cite{Miszczak10} for twice as many qubits as there are
system qubits. The random matrices were hermitized, multiplied by a 
randomly chosen scaling factor and exponentiated. The 
resulting matrix was multiplied by the tensor product of the target
unitary with $\openone$, and the bath qubits were traced out.
For most dynamical maps, $F^{arith}_{unitary}$
yields a good estimate of the gate error. 
If, however, the state fidelities for all $\rho_{B,i}$ are very
high, but the fidelity for the totally rotated state is comparatively
small, the arithmetic mean seriously underestimates the gate error.
This can happen, for example, if the evolution is perfectly 
unitary, $\mathcal D(\rho_j)=\tilde U\rho_j\tilde U^+$, and $\tilde U$
and the target $O$ have a common eigenbasis with all the 
$\rho_{B,i}$. Then the information relevant for the gate error is 
completely contained in 
$F_{TR}$. This is illustrated in Fig.~\ref{fig:histo}(e) for
randomized unitaries with an eigenbasis very close to the
$\rho_{B,i}$ and $O=\openone$. In such a case the geometric average 
over all state fidelities 
will yield a much better estimate of the gate fidelity. 
In most cases, however, the geometric mean is too strict and
overestimates the gate error, motivating the
definition~\eqref{eq:Fred}. Indeed, the best estimates of the gate
error are obtained using
$\varepsilon_{unitary}^{\lambda}=1-F^{\lambda}_{unitary}$
as shown in the right part of Fig.~\ref{fig:histo}. 
Figure~\ref{fig:histo}(a,b,e,f) presents results for randomized
dynamical maps and 
randomized unitaries that were generated by exponentiating random
Hermitian matrices. Since this is not truly random,
we have also generated random unitaries based on
Gram-Schmidt orthonormalization of randomly generated complex
matrices~\cite{MezzadriNAMS07}, cf. Fig.~\ref{fig:histo}(c,d) with 
$O=$CNOT. $\varepsilon_{unitary}^{\lambda}$ yields a faithful estimate
of the gate error in all cases. On average, it
underestimates the gate error by factors 1.03
(Fig.~\ref{fig:histo}b), 1.11 (d) and 1.02 (f)
and overestimates it by 1.16 (b),
1.08 (d), and 1.01 (f). 
This illustrates that $F_{unitary}^{\lambda}$ makes best use of the
information contained in the $d+1$ state fidelities, $F_{B,i}$
and $F_{TR}$. 
\begin{table}[tb]
  \centering
  \begin{tabular}{|c|l|c|c|c|c|}
    \hline
    $N$ & type of dynamics & $\alpha^{arith}$ & $\beta^{arith}$ & $\alpha^{\lambda}$ &
    $\beta^{\lambda}$ \\ \hline
    2 & randomized dynamical map & 0.83 & 1.31 & 0.44 & 1.26 \\
    & random unitaries & 0.76 & 2.35 & 0.75 & 1.92 \\
    & randomized unitaries & 1.00 & 4.39 & 0.90 & 1.15\\ \hline
    3 & randomized dynamical map & 0.96 & 1.04 & 0.51 & 1.03 \\
    & random unitaries & 0.90 & 1.32 & 0.90 & 1.32 \\
    & randomized unitaries & 1.00 & 8.67 & 0.91 & 1.20 \\ \hline    
  \end{tabular}
  \caption{Numerically obtained bounds for over- and underestimation
    of the average 
    fidelity of the form $\alpha^i F^i_{unitary} \le F_{av} \le \beta^i
    F^i_{unitary}$ for the arithmetic mean over the state fidelities
    ($i=arith$) and the combination of arithmetic and geometric mean
    ($i=\lambda$), using 
    100.000 realizations, for 2 and 3 qubits with $O$ corresponding to
    CNOT ($N=2$), the Toffoli gate ($N=3$) and identity
    (randomized unitaries).} 
  \label{tab:bounds}
\end{table}
Bounds for over- and underestimating the gate error, 
obtained numerically, 
are presented in 
Table~\ref{tab:bounds} with CNOT, identity and the Toffoli gate as
target operations.  
For three-qubit gates, we find the numerical bounds to be essentially
contained by those for two-qubit gates,
cf. Table~\ref{tab:bounds}. This suggests 
our numerical bounds to be independent of system size. A verification
of this conjecture for larger system sizes is, however, hampered by
the enormous increase in numerical effort for randomization. For our
examples of CNOT, the Toffoli gate and identity, we find the estimated
gate error based on Eqs.~\eqref{eq:Fred} to 
deviate from the standard one in the worst case 
by a factor smaller than 2.5 and on average by a factor smaller than
1.2. This confirms that  $d+1$ state fidelities $F_j$ are 
sufficient to accurately estimate the gate error.

\subsection{Quantifying non-unitarity}
\label{subsec:nonunitarity}
If, in a given experimental setting, the gate error turns out to be
larger than expected, one might want to know whether it is due to
unitary errors or 
decoherence. This can be determined by quantifying 
non-unitarity of the time evolution using 
the following distance measure,
\begin{equation}
  \label{eq:Fdiss}
  F_{diss} = 1-\frac{1}{d+1} \left\{
    \sum_{i=1}^d\Tr\left[\rho^2_{B,i}(T)\right] 
    + \Tr\left[\rho^2_{TR}(T)\right]
  \right\}\,,
\end{equation}
where $\rho_j(T)=\mathcal D(\rho_j)$. $F_{diss}=0$ if and only if the
evolution is completely unitary. Evaluation of $F_{diss}$ 
requires preparation of the $d+1$ input states of Eq.~\eqref{eq:rho}
and measurement of $d^2+d$ populations. 

Equation~\eqref{eq:Fdiss} cannot replace full process tomography when
complete identification of the error sources is desired. However, some 
information can already be gained by inspection of the $d+1$ purities
of Eq.~\eqref{eq:Fdiss}. For example, if the purity loss is due to a
single or very few terms in Eq.~\eqref{eq:Fdiss}, this identifies the
state evolutions that are subject to dissipation. On the other hand,
if the purity loss is equally distributed over all basis states, the
chosen basis is likely not  an eigenbasis of the error 
operators (but another mutually unbiased basis presumably is).

\section{Reduced set of states yielding analytical bounds on the 
  average fidelity} 
\label{sec:Hofmann}

We now connect our notion of a reduced set of input states to the
result of Ref.~\cite{HofmannPRL05} that two classical fidelities
can be used to obtain an upper and a lower bound on the average fidelity. 
The classical fidelity is given by the average
probability of obtaining the correct output for each of the $d$ classically
possible input states,
\begin{eqnarray}
F_{c} & = & \frac{1}{N}\sum_{i=1}^{d}
\langle k_{i}^{\left(1\right)}|U_{0}^{\dagger}\mathcal{D}(|k_{i}^{\left(1\right)}\rangle
\langle k_{i}^{\left(1\right)}|)U_{0}|k_{i}^{\left(1\right)}\rangle
\end{eqnarray}
for an arbitrary orthonormal Hilbert space basis 
$\{|k_{i}^{\left(1\right)}\rangle \}_{i=1,\dots,d}$.
It can be interpreted as the arithmetic average over the overlaps
between expected and actual population evoluation for the basis states
$\ket{k_{i}^{\left(1\right)}}$. Defining 
$\rho_{i}^{\left(1\right)}=|k_{i}^{\left(1\right)}\rangle\langle k_{i}^{\left(1\right)}|$,
such a classical fidelity can be rewritten analogously to
Eq.~\eqref{eq:Farith}, 
\begin{eqnarray*}
  F_{1}  & = &
  \frac{1}{N}\sum_{i=1}^{d}\langle k_{i}^{\left(1\right)}|
  U_{0}\mathcal{D}\left(\rho_{i}^{\left(1\right)}\right)
  U_{0}^{\dagger}|k_{i}^{\left(1\right)}\rangle\\
   &=&  \frac{1}{N}\sum_{ij=1}^{d}\langle k_{j}^{\left(1\right)}|k_{i}^{\left(1\right)}\rangle
  \langle k_{i}^{\left(1\right)}|U_{0}\mathcal{D}\left(\rho_{i}^{\left(1\right)}\right)
  U_{0}^{\dagger}|k_{j}^{\left(1\right)}\rangle\\
  & = & \frac{1}{N}\sum_{i=1}^{d}\text{Tr}
 \left[\rho_{i}^{\left(1\right)}U_{0}\mathcal{D}\left(\rho_{i}^{\left(1\right)}\right)
   U_{0}^{\dagger}\right] \\ &=& 
 \frac{1}{N}\sum_{i=1}^{d}\text{Tr}
 \left[U_{0}^{\dagger}\rho_{i}^{\left(1\right)}U_{0}
 \mathcal{D}\left(\rho_{i}^{\left(1\right)}\right)\right]\,,
\end{eqnarray*}
with $U_{0}^{\dagger}\rho_{i}^{\left(1\right)}U_{0}$ the ideal and 
$ \mathcal{D}\left(\rho_{i}^{\left(1\right)}\right)$
the actual evolutions.
In our terminology, the states $\rho_i^{(1)}$ 'fix' the basis,
cf. Eq.~\eqref{eq:rho_Bi}. In order to fulfill the requirements of a
reduced set, i.e., to allow for differentiating any two unitaries and
assessing unitarity of the time evolution, another state corresponding
to the totally rotated projector is necessary,
cf. Section~\ref{sec:algframe}. Instead of a single state 
$\rho_{TR}$, Ref.~\cite{HofmannPRL05} chooses $d$ such states with
each state fulfilling the condition of total rotation:
$\rho_{i}^{\left(2\right)}=|k_{i}^{\left(2\right)}\rangle\langle
k_{i}^{\left(2\right)}|$ with 
\begin{equation*}
\left|\Braket{k_{i}^{\left(1\right)}|k_{j}^{\left(2\right)}}\right|^{2}=\frac{1}{d}\quad
\forall i,j\,,
\end{equation*}
i.e., a complete mutually unbiased basis~\cite{FernandezPRA11}.
Evaluating the classical fidelities for the two bases
$\{|k_{i}^{\left(1\right)}\rangle \}_{i=1,\dots,d}$, 
$\{|k_{i}^{\left(2\right)}\rangle \}_{i=1,\dots,d}$ then allows for
analytical bounds on the average fidelity~\cite{HofmannPRL05}. 

Note that Ref.~\cite{HofmannPRL05} discusses a specific choice of the
two bases. From our derivation in Section~\ref{sec:algframe}, it is
clear that any two mutually unbiased bases are suitable, and  one
can choose the most convenient ones. There exist
$d+1=2^N+1$ mutually unbiased bases for $N$
qubits~\cite{WoottersAnnPhys89}, but only three of them consist of
separable states while the remaining $d-2$ mutually unbiased bases are
made up of maximally entangled states~\cite{LawrencePRA11}. 
Any two of the three separable mutually unbiased bases 
constitute a natural choice for most experimental setups.

\section{Conclusions}
\label{sec:concl}
We have shown that a reduced set of input states can be used to
estimate the average fidelity  
or gate error of a quantum gate. It provides  the
information to characterize, instead of the full open
system evolution, only the unitary part.
The average over all Hilbert space can then be estimated by a
modified average over a reduced set of states that allows to differentiate any two unitaries
and quantify non-unitarity of the evolution. The states in the 
reduced set correspond to a complete set of orthonormal
one-dimensional projectors plus a one-dimensional
projector that is rotated with respect to all the other
projectors. The reduced set can be realized by two
mixed states, irrespective of the dimension $d$ of the
Hilbert space, or by $d+1$ pure states. 
Our concept of total rotation 
is related to the notion of mutually unbiased
bases~\cite{FernandezPRA11} where all states of 
the second basis are totally rotated with respect to the first basis. 
It is also the underlying principle in constructing the 
input states for two complementary classical
fidelities~\cite{HofmannPRL05}. Consequently, one can estimate the
average fidelity using $d+1$ or $2d$ pure separable input states. 
In both cases, the gate error is determined in terms of 
state fidelities for the time-evolved states of the reduced
set. The approach using $2d$ input states comes with the advantage of
analytical bounds on the gate error. For the 
smaller set of $d+1$ input states, numerical calculations demonstrate the
estimate to deviate from the true gate error  by less than a
factor 1.2 on average and less than a factor 2.5 in the worst case.

While the average gate fidelity currently enjoys great popularity,
other 
very useful performance measures exist~\cite{GilchristPRA05}. For
example, the worst case fidelity is relevant in the context of the
error correction threshold~\cite{BenyPRL10}. It would be interesting
to see whether the $d+1$ or $2d$ state fidelities of the reduced set
allow for estimating bounds on the worst-case fidelity. This is,
however, beyond the scope of our current work.

Another important question concerns the scaling of the gate error estimate
employing a reduced set of states with the number of qubits. The 
straightforward, but not most efficient approach consists in
determining the required $d+1$ or $2d$ state fidelities by 
state tomography. This yields a scaling of $2^{3N}$ for standard state
tomography and $2^{2N}$ for Monte Carlo state
characterization~\cite{FlammiaPRL11,daSilvaPRL11}, i.e., no
improvement over current approaches. Alternatively, a reduced set
of states can be combined with Monte Carlo process
characterization~\cite{FlammiaPRL11,daSilvaPRL11}. We show in
Ref.~\cite{Reich13b} that in this case the experimental effort and the
classical computational resources to obtain tight analytical bounds on
the average error are reduced by a factor $2^N$
compared to the best currently available protocol for
general unitary operations. 
 
The ability to measure the gate performance efficiently with a reduced
set of input states is not only a
prerequisite for the development of quantum devices; it also opens the
door to designing quantum gates in coherent control experiments 
using e.g. genetic algorithms where repeated checks of the performance
are required. 

\begin{acknowledgments}
  We would like to thank Christian Roos for helpful comments. 
  Financial support from the EC
  through the FP7-People IEF Marie Curie action Grant
  No. PIEF-GA-2009-254174 is gratefully acknowledged.
\end{acknowledgments}

\appendix

\section{Proofs}
\label{sec:app}
We provide here detailed proofs of the claims made in
Section~\ref{sec:algframe}.

\subsection{Injectivity of $\mathcal M$ is equivalent to the
  commutant space $\mathcal K(\{\rho_i\})$ having identity as its only
  element}

\textbf{Definition:}
Let $\rho$ be a density operator defined in a $d$-dimensional Hilbert
space  and $U_i$ elements of the projective unitary group $PU(d)$. 
We call the set of operators
\[
K\left(\rho\right) = \left\{ U_{i}\in PU(d) \,|\,
\left[U_{i},\rho\right]=0\right\} 
\]
the commutant space of $\rho$ in $PU(d)$. The commutant space of a set
of density operators $\left\{ \rho_{j}\right\}$  is defined as
\[
K\left(\{\rho_j\}\right)=\bigcap_{j}K\left(\rho_{j}\right)\,.
\]

\textbf{Proposition:}
The map $\mathcal{M}: PU(d) \to \bigoplus_{i}\mathbb{C}^{n\times n}$ 
which maps any unitary $U\in\text{PU\ensuremath{\left(d\right)}}$ 
to the set of propagated density operators 
$\left\{ \rho_{i}^{(U)}\left(T\right)\right\} $ is injective 
iff the commutant space of $\{\rho_i\}$ in $PU(d)$, $\mathcal
K(\{\rho_i\})$,  contains only the
identity. 

\textbf{Proof:}
Injectivity of $\mathcal{M}$ is equivalent to the condition
\[
\forall i:\rho_{i}^{\left(U\right)}\left(T\right)=
\rho_{i}^{\left(V\right)}\left(T\right)\Longleftrightarrow U=V
\]
We first show that this condition is equivalent to 
\[
\forall i:\rho_{i}^{\left(U\right)}\left(T\right)=\rho_{i}
\Longleftrightarrow U=\openone
\] 
Assuming validity of 
$\forall i:\rho_{i}^{\left(U\right)}\left(T\right)=
\rho_{i}^{\left(V\right)}\left(T\right)\Longleftrightarrow U=V$
just choose $V=\openone$. Then
$\rho_{i}^{\left(V\right)}\left(T\right)=\rho_{i}$ 
and the second statement follows immediately. 
Conversely, assume 
\[
\forall i:\rho_{i}^{\left(U\right)}\left(T\right)=\rho_{i}
\Longleftrightarrow U=\openone
\]
Then, 
for arbitrary $V,W\in PU(d)$ we set
$U=V^{-1}W=V^{+}W$ and 
\begin{eqnarray*}
\forall i:\rho_{i}^{\left(V^{+}W\right)}\left(T\right) =  \rho_{i}
&\Longleftrightarrow &\forall i:V^{+}W\rho_{i}W^{+}V  =  \rho_{i}\\
&\Longleftrightarrow&\forall i:W\rho_{i}W^{+}  =  V\rho_{i}V^{+}\\
&\Longleftrightarrow&\forall i:\rho_{i}^{\left(W\right)}\left(T\right) 
= \rho_{i}^{\left(V\right)}\left(T\right)
\end{eqnarray*}
By assumption 
$\forall i:\rho_{i}^{\left(V^{+}W\right)}\left(T\right)=\rho_{i}
\Longleftrightarrow V^{+}W=\openone$,
but since 
\[
\forall i:\rho_{i}^{\left(V^{+}W\right)}\left(T\right)=\rho_{i}
\Longleftrightarrow
\forall
i:\rho_{i}^{\left(W\right)}\left(T\right)=\rho_{i}^{\left(V\right)}\left(T\right)
\]
and the relation 
\[
V^{+}W=\openone\Longleftrightarrow W=V
\] 
always holds for $V,W\in PU(d)$, this 
leads to the desired result: 
$\forall i:\rho_{i}^{\left(U\right)}\left(T\right)=
\rho_{i}^{\left(V\right)}\left(T\right)\Longleftrightarrow U=V$.
 
We now show that 
$\mathcal K(\{ \rho_{i}\} )=\openone$
iff 
\[
\forall i:\rho_{i}^{\left(U\right)}\left(T\right)=
\rho_{i}\Longleftrightarrow U=\openone
\]
Consider the following calculation 
\begin{eqnarray*}
\forall i:\rho_{i}^{\left(U\right)}\left(T\right)  = \rho_{i}
&\Longleftrightarrow&\forall i:U\rho_{i}U^{+} = \rho_{i}\\
&\Longleftrightarrow&\forall i:U\rho_{i} =  \rho_{i}U\\
&\Longleftrightarrow&\forall i:U\rho_{i}-\rho_{i}U  =  0\\
&\Longleftrightarrow&\forall i:\left[U,\rho_{i}\right]  =  0\\
&\overset{\left(*\right)}{\Longleftrightarrow}&U  =  \openone
\end{eqnarray*}
The equivalence relation $\left(*\right)$ is true only iff
$\mathcal K(\{ \rho_{i}\} )=\openone$.
This concludes the proof.

\subsection{Total rotation and commutation with identity}

\textbf{Definition:}
Let $\mathcal{H}$ be a $d$-dimensional Hilbert space. A set $\mathcal
P_c$ of $d$ one-dimensional orthogonal projectors from $\mathcal{H}$
onto itself is called complete. For example, spanning the Hilbert
space by an arbitrary complete orthonormal basis $\{\ket{\varphi_i}|\}$, $\mathcal
P_c=\{P_i=\ket{\varphi_i}\bra{\varphi_i}\}$. A one-dimensional projector
$P_{TR}$ from $\mathcal{H}$
onto itself is called totally rotated with respect to the set $\mathcal{P}_c$
if $\forall P_i\in\mathcal{P}_c:\ P_{TR}P_i\neq 0$.
A set $\mathcal{P}_{cTR}\equiv\left\{ \mathcal P_c, P_{TR}\right\}$
of projectors is called complete and totally rotating.

\textbf{Definition:}
Let $\mathcal{H}$ be a $d$-dimensional Hilbert space. A set of density
operators $\left\{ \rho_{i}\right\}$ 
with $\rho_{i}\in\mathcal{H}\otimes\mathcal{H}$ is called complete
if the set of projectors on the eigenspaces of the $\left\{ \rho_{i}\right\} $
is complete and 
is called complete and totally rotating if the set of projectors on
the eigenspaces of the $\left\{ \rho_{i}\right\} $ is complete and
totally rotating.

Our goal is to prove that the only projective unitary matrix that
commutes with each element of a complete and totally rotating set of
states is the identity. 
In order to make use of the assumed commutation relations in the
proof, we translate commutation of a unitary with a state into
commutation of a unitary with one or more projectors. To this end, we
introduce a lemma. Using commutation of a unitary with projectors, it
is then straightforward to show that the unitary must be the identity.

\textbf{Lemma:}
If $\left[U,\rho\right]=0$ for a unitary 
$U \in PU(d)$ and a density operator $\rho$ which has at least one 
non-degenerate eigenvalue $\lambda_{1}$, then
$\left[U,P_{1}\right]=0$ where $P_{1}$
is the projector onto the eigenspace $\mathcal E_1$
corresponding to the eigenvalue $\lambda_1$.

\textbf{Proof:}
Since $\rho$ has a non-degenerate eigenvalue $\lambda_{1}$,
we can expand it in a set of orthonormal projectors,
$\rho=\lambda_{1}P_{1}+\sum_{i=2}^d\lambda_{i}P_{i}$, 
with $P_{1}=\ket{\xi_1}\bra{\xi_1}$ the projector onto the
one-dimensional eigenspace $\mathcal E_1$.
By assumption,
\begin{eqnarray*}
  \left[U,\rho\right] = 0
  &=& \lambda_{1}UP_{1}-\lambda_{1}P_{1}U +
  \sum_{i=2}^d\left(\lambda_{i}UP_{i}-\lambda_{i}P_{i}U\right)  \\
  &=& \lambda_{1}UP_{1}U^+-\lambda_{1}P_{1} + 
  \sum_{i=2}^d\left(\lambda_{i}UP_{i}U^+-\lambda_{i}P_{i}\right)\,,
\end{eqnarray*}
where in the second line we have multiplied by $U^+$ from the right.
Defining $\bar P_{i}=UP_{i}U^+$,  
this is equivalent to
\[
\lambda_{1}\bar P_{1}+\sum_{i=2}^d\lambda_{i}\bar P_{i}
= \lambda_{1}P_{1}+\sum_{i=2}^d\lambda_{i}P_{i}\,.
\]
The operator equality can be applied to $\ket{\xi_1}$, leading to 
\[
\lambda_{1}\bar{P}_{1}\ket{\xi_1} + 
\sum_{i=2}^d\lambda_{i}\bar{P}_{i}\ket{\xi_1} = \lambda_{1}\ket{\xi_1}\,.
\]
Inserting identity, $\sum_{i=1}^d\bar{P}_{i}=\openone$, in the
right-hand side, we obtain
\[
\lambda_{1}\bar{P}_{1}\ket{\xi_1} + 
\sum_{i=2}^d\lambda_{i}\bar{P}_{i}\ket{\xi_1} =
\lambda_{1}\bar{P}_{1}\ket{\xi_1}+\sum_{i=2}^d\lambda_{1}\bar{P}_{i}\ket{\xi_1}\,. 
\]
Multiplying from the left by $\bar P_{i\neq 1}$ and using
orthogonality of the $\bar P_i$ and non-degeneracy of $\lambda_1$,
$\lambda_{i\neq 1}\neq \lambda_1$, we find that $\bar
P_i\ket{\xi_1}=0$ for all $i\neq 1$. Therefore $\ket{\xi_1}$
lies also in the one-dimensional eigenspace corresponding to $\bar
P_1$, 
and the one-dimensional eigenspaces of $P_1$ and $\bar P_1$ must be 
identical. This implies 
\[
P_1 = \bar P_1\,,
\]
and, by definition of $\bar P_1$, we find that $U$
leaves the one-dimensional eigenspace corresponding to $P_1$
invariant, hence commutes with $P_1$.

Note that if a density operator $\rho$ that commutes with $U$ has more
than one non-degenerate eigenvalue, the lemma implies commutation of
$U$ with all the projectors onto the one-dimensional eigenspaces.

\textbf{Proposition:}
The only projective unitary matrix that commutes with a set of states
$\{\rho_i\}$ that is complete and totally rotating is the identity.

\textbf{Proof:} 
Repeated application of the lemma to states $\rho_i$
yields a set of one-dimensional projectors that each commute with $U$. By
definition of a complete and totally rotating set of states, 
$d+1$ projectors within this set 
must be elements of $\mathcal \{\mathcal P_c,P_{TR}\}$.
We can thus choose the complete set of one-dimensional projectors 
$\mathcal P_c$ to represent $U$, $U=\sum_{i=1}^du_i P_i$. 
An equally valid choice $\{\tilde P_i\}$ employs the totally rotated
projector, $\tilde P_1 = P_{TR}$, with $\mathcal E_{TR}$ the
corresponding eigenspace, 
and a suitable set of orthonormal one-dimensional projectors $\{\tilde
P_i\}_{i=2,\dots,d}$ 
for the space $\mathcal E_{TR}^\bot$ such that $U=\sum_{i=1}^du_i
\tilde P_i$. The spectrum $\{u_i\}$ is of course independent of the
representation.  
Consider the action of $U$ on a vector $\ket{\zeta} \in \mathcal 
E_{TR}$, 
\begin{eqnarray}\nonumber
  U\ket{\zeta} &=&\sum_{i=1}^du_{i}P_{i}\ket{\zeta} \\ &=&
  \sum_{i=1}^du_{i}\tilde{P}_{i}\ket{\zeta} = 
  u_{1}\ket{\zeta} = \sum_{i=1}^du_{1}P_{i}\ket{\zeta}\,, 
  \label{eq:therorem715}
\end{eqnarray}
where we have inserted $\sum_{i=1}^dP_{i}=\openone$ in the last step. 
By total rotation, $\ P_{TR}P_{i}\neq0$ $\forall
P_i\in\mathcal P_c$, or equivalently, $P_{i}P_{TR}\neq0$.
Applying this to $\ket{\zeta}$, we find
\[
\ P_{i}P_{TR}\ket{\zeta}=P_{i}\ket{\zeta}\neq 0 \quad \forall i\,.
\]
Since the $P_{i}$ are one-dimensional orthonormal projectors, 
i.e., $P_i=\ket{\varphi_i}\bra{\varphi_i}$ with
$\{\ket{\varphi_i}\}$ a complete orthonormal basis
of the Hilbert space, we can rewrite $P_{i}\ket{\zeta}$,
\[
  P_{i}\ket{\zeta}=\mu_{i}\ket{\varphi_i}   
\]
with $\mu_{i}\in\mathbb{C}$, $\mu_{i}\neq0$. 
Inserting this into Eq.~\eqref{eq:therorem715}, we obtain
\[
\sum_{i=1}^du_{1}\mu_{i}\ket{\varphi_i}=\sum_{i=1}^du_{i}\mu_{i}\ket{\varphi_i}\,. 
\]
Comparing the coefficients yields 
$u_{1}\mu_{i}=u_{i}\mu_{i}$ $\forall i$. 
Since all $\mu_{i}\neq0$ due to total rotation, we can divide and
obtain
\[
u_{1}=u_{i} \quad \forall i \,,
\]
i.e., a unitary with complete degeneracy in its eigenvalues.
This necessarily has to be the matrix $e^{i\varphi}\openone$ for
$\varphi\in\left[0,2\pi\right]$,
or, as an element of $PU(d)$, the unit matrix.

We have thus shown that only the identity commutes with a set of
states that is complete and totally rotating. This set of states is
therefore sufficient to differentiate any two unitaries. 

\subsection{Unitarity of $\mathcal{D}$ is equivalent to projectors
  being mapped onto projectors}

\textbf{Proposition:}
A dynamical map $\mathcal D$, defined on a $d$-dimensional Hilbert
space, is unitary if and only if $\mathcal D$ maps (i) a set $\{P_i\}$ of
$d$ one-dimensional orthonormal projectors onto a set of
$d$ one-dimensional orthonormal projectors $\{\tilde P_i\}$ and (ii) a
one-dimensional projector that is totally rotated with respect to 
$\{P_i\}$ onto a one-dimensional projector (which is totally rotated
with respect to $\{\tilde P_i\}$). 

\textbf{Proof:}
We first prove the forward direction.
If the time evolution is unitary, the action of $\mathcal D$ on any
state is described by $\mathcal D(\rho)=U\rho U^+$. Specifically for
orthonormal projectors $P_iP_j=\delta_{ij}$, we find 
\[
\mathcal{D}\left(P_{i}\right)\mathcal{D}\left(P_{j}\right) 
= UP_{i}U^+UP_{j}U^{\dagger}
= UP_{i}P_{j}U^{\dagger}=\delta_{ij}UP_{i}U^{\dagger}\,.
\]
Since a one-dimensional projector can be written 
$P_i=|\varphi_i\rangle\langle \varphi_i|$, where $\{\ket{\varphi_i}\}$
is a complete orthornormal basis of $\mathcal H$, 
$UP_{i}U^{\dagger}$ is also one-dimensional projector.
By the same argument, $P_{TR}$ is mapped onto a one-dimensional
projector if $\mathcal D(\rho)=U\rho U^+$.
Therefore 
a dynamical map $\mathcal D$ describing unitary time evolution maps
a set of $d$ orthonormal projectors, $\{P_i\}$, 
onto another such set, $\{\tilde P_i = UP_{i}U^+\}$, and $P_{TR}$ onto
a one-dimensional projector. 

We now prove the backward direction, starting from the representation
of $\mathcal D$, 
\begin{equation}
  \label{eq:defD}
  \mathcal D = \sum_{k=1}^K E_k \rho E_k^+\,,
\end{equation}
by Kraus operators $E_k$, i.e., linear operators that
fulfill 
\begin{equation}
  \label{eq:Ecomplete}
  \sum_{k=1}^KE_{k}^+E_{k}=\openone\,. 
\end{equation}
We employ the canonical representation in which the Kraus operators are
orthogonal, $\Tr\left[E^+_k E_l\right] \sim \delta_{kl}$. By
assumption, a set of $d$ one-dimensional, orthonormal
projectors $\{P_{i}\}$ is mapped by $\mathcal D$ onto another such set
$\{\tilde{P}_{i}\}$, 
\begin{equation}
\label{eq:mapP}
\mathcal{D}\left(P_{i}\right)=\sum_{k=1}^KE_{k}P_{i}E_{k}^+=\tilde{P}_{i}\,. 
\end{equation}
We need to show that this implies $\mathcal D(\rho)=U\rho U^+$, or
equivalently, as we demonstrate below, that $\mathcal D$ is made up of
a single Kraus operator $E_1$ in the representation
where $\Tr\left[E^+_k E_l\right] \sim \delta_{kl}$. In general, 
we can employ a polar decomposition for each Kraus operator,
factorizing it into a
unitary and a positive-semidefinite operator, 
$E_k=U_k\tilde E_k$. For unitary evolution, $U_k=\tilde U$ for all $k$
and $E_1=U\openone$ which is a special case of $\tilde E_k$
being diagonal. We first show that the assumption for the $d$
orhonormal projectors $\{P_i\}$ implies $U_k=\tilde U$ and 
diagonality of $\tilde E_k$.
In a second step, we prove that the assumption for the totally
rotated projector implies that there is only a single Kraus operator
and $\tilde E_1=\openone$. 

We first show that $\tilde E_k=E_k U^+$ is diagonal in the orthonormal
basis $\{\ket{\varphi_i}\}$ corresponding to the $P_i$.
Equation~\eqref{eq:mapP} suggests the definition of an 
operator $\Pi_{k}^{\left(i\right)}\equiv E_{k}P_{i}E_{k}^+$ which is
obviously Hermitian and moreover semipositive definite.
The latter is seen by making use of
$P_i^2=P_i$ and $P_i=P_i^+$:
$\left\langle \zeta\big|\Pi_{k}^{\left(i\right)}\zeta\right\rangle =
\left\langle \zeta|E_{k}P_{i}P_{i}E_{k}^+\zeta\right\rangle =
\left\langle P_{i}E_{k}^+\zeta|P_{i}E_{k}^+\zeta\right\rangle =
\left\langle \xi|\xi\right\rangle \geq 0$ for any
$\ket{\zeta}\in \mathcal H$.
Equation~\eqref{eq:mapP} implies 
$\sum_{k=1}^K\Pi_{k}^{\left(i\right)}=\tilde{P}_{i}$. For the 
normalized vector spanning the eigenspace of $\tilde{P}_{i}$,
$\ket{\tilde\varphi_i} \in \mathcal E_i$, we find 
\[
\sum_{k=1}^K\left\langle \tilde\varphi_i\bigg|
\Pi_{k}^{\left(i\right)}\tilde\varphi_i\right\rangle =1\,,
\]
while for all $\ket{\xi} \in \mathcal E_i^\bot$ 
\[
\sum_{k=1}^K\left\langle \xi\bigg|
\Pi_{k}^{\left(i\right)}\xi\right\rangle =0\,.
\]
Due to positive semidefiniteness of $\Pi_{k}^{\left(i\right)}$, this
implies 
$\left\langle \xi\big|
\Pi_{k}^{\left(i\right)}\xi\right\rangle =0$.
Reinserting the definition of $\Pi_{k}^{\left(i\right)}$ leads to
$\left\langle \xi|
E_{k}P_{i}E_{k}^+\xi\right\rangle =
\left\langle P_{i}E_{k}^+\xi|
P_{i}E_{k}^+\xi\right\rangle=0$, i.e., we find  
$P_{i}E_{k}^+\ket{\xi}=0$ for all 
$k$, $i$ and $\ket{\xi}\in \mathcal E_{i}^\bot$. 
For an arbitrary Hilbert space vector $\ket{\zeta}$,
$\left(\openone-\tilde P_i\right)\ket{\zeta}$ lies in  $\mathcal
E_{i}^\bot$ such that 
$P_{i}E_{k}^+\left(\openone-\tilde P_i\right)\ket{\zeta}=0$ for all 
$k$ and $i$. Therefore 
\[
P_{i}E_{k}^+\left(\openone-\tilde{P}_{i}\right) =  0\,, 
\quad\mathrm{or}\,,\quad
P_{i}E_{k}^+ = P_{i}E_{k}^+\tilde{P}_{i} \quad \forall i,k\,.
\]
To make use of the orthogonality of the $\tilde P_i$, we
multiply by $\tilde P_j$, $j\neq i$ from the right. Since $\tilde P_j$
can be written as $\tilde{P}_{j}=\tilde UP_{j}\tilde U^+$ for a
specific $\tilde U$, 
we obtain, for all $i$, $k$ and
$j\neq i$, 
$P_{i}E_{k}^+ \tilde U P_{j}\tilde U^+=0$.
Multiplication by $\tilde U$ from the right yields
\[
P_{i}E_{k}^+ \tilde UP_{j} =0\,.
\]
This implies that the operators $E_{k}^+ \tilde U$ have to be
diagonal in the basis corresponding to the $P_i$, 
\begin{equation}
  \label{eq:Ekdiag}
E_{k}^+ \tilde U = \sum_{i=1}^d e^k_i P_i\,.  
\end{equation}
Note that the unitary $\tilde U$ is the same for all Kraus operators
$E_k$. 

In the second step, we now need to show that the right-hand side of
Eq.~\eqref{eq:Ekdiag} is equal to the identity, making use of the
assumption that the totally rotated projector is mapped by $\mathcal D$ onto a
one-dimensional projector. The crucial information is captured in the
coefficients $e^k_i$. Let us summarize what we know about the
$e^k_i$. From orthogonality of the Kraus operators, we find 
$\Tr\left[E_{k}^+E_{l}\right] =
\Tr\left[\sum_{i,j=1}^de^k_{i}(e^l_{j})^{*}P_{i}\tilde U^+\tilde UP_{j}\right] =
\sum_{i,j=1}^de^k_{i}(e^l_{j})^{*}\Tr\left[P_{i}P_{j}\right]=
\sum_{i,j=1}^de^k_{i}(e^l_{j})^{*}\delta_{ij} = 
\sum_{i=1}^de^k_{i}(e^l_{i})^{*}\overset{!}{\sim}\delta_{kl}$. The
last sum can be interpreted as a scalar product for two orthogonal 
vectors $\vec{e}^{\;k},\vec{e}^{\;l} \in \mathbb{C}^{d}$ with
coefficients $e^k_{i}$, $e^l_{i}$. Defining the proportionality
constants $\mathcal N(k)$, 
\begin{equation}
  \label{eq:defN}
\mathcal{N}\left(k\right) \equiv \Tr\left[E_{k}^+E_{k}\right]
=\sum_{i}e^k_{i}(e^k_{i})^{*} = \langle
\vec{e}^{\;k},\vec{e}^{\;k}\rangle \ge 0\,,  
\end{equation}
we find from Eq.~\eqref{eq:Ecomplete} and $\Tr[\openone]=d$ that
$\sum_{k=1}^K\mathcal N(k) = d$
(and, if we can show that $\mathcal N(k)=d$ for one $k$, than the
number of Kraus operators, $K$, must be one). 
Equation~\eqref{eq:Ecomplete} together with Eq.~\eqref{eq:Ekdiag}
yields yet another condition on the $e^k_i$: 
$\openone =\sum_{k=1}^KE_{k}^+E_{k} = 
\sum_{i,j=1}^d\sum_{k=1}^K(e^k_{i})^{*}e^k_{j}P_{i}P_{j} =
\sum_{i,k}\left|e^k_{i}\right|^{2}P_{i}$ such that 
$\sum_{k}\left|e^k_{i}\right|^{2}=1$ for
each $i$. This can be interpreted as normalization condition for a
vector $\vec{\epsilon}_i\in \mathbb{C}^{K}$ with coefficients
$e^k_i$,
\begin{equation}
  \label{eq:normeps}
  1 = \sum_{k=1}^K\left|e^k_{i}\right|^{2} = 
  \langle \vec{\epsilon}_i,\vec{\epsilon}_i\rangle\,.  
\end{equation}
Since the vector sets $\{\vec{e}_k\}$ and $\{\vec{\epsilon}_i\}$ are not
independent, it is clear that any information on the scalar product 
$\langle \vec{\epsilon}_i,\vec{\epsilon}_j\rangle$ will be useful to
determine $\mathcal N(k)$ (such that we can check whether there is one 
$k$ for which $\mathcal N(k)=d$).  To this end, we employ the
assumption that $P_{TR}$ is mapped by $\mathcal D$ onto a
one-dimensional projector, $\tilde P_{TR}=\mathcal D(P_{TR})$, or, in
other words the purity of  $P_{TR}$ is preserved, 
\[
\Tr\left[\left(\mathcal D(P_{TR})\right)^2\right]=1\,.
\]
Inserting Eqs.~\eqref{eq:defD} and \eqref{eq:Ekdiag}, making use of
the orthogonality of the $P_i$ and of the trace being invariant under
cyclic permutation, we find
\begin{widetext}
\begin{eqnarray*}
  \Tr\left[\mathcal{D}\left(P_{TR}\right)^{2}\right] & = & 
  \Tr\left[U\left(\sum_{ij}\sum_{k}\sum_{i^{'}j^{'}}\sum_{k^{'}}
      (e^k_{i})^{*}e^k_{j}(e^{k^{'}}_{i^{'}})^{*}e^{k^{'}}_{j^{'}}
      P_{i}P_{TR}P_{j}P_{i^{'}}P_{TR}P_{j^{'}}\right)U^+\right]\\
  & = & \Tr\left[\sum_{ij}\sum_{k}\sum_{j^{'}}\sum_{k^{'}}
    (e^k_{i})^{*}e^k_{j}(e^{k^{'}}_{j})^{*}e^{k^{'}}_{j^{'}}
    P_{i}P_{TR}P_{j}P_{TR}P_{j^{'}}\right]\\
  & = & \sum_{ij}\sum_{k}\sum_{j^{'}}\sum_{k^{'}}
  (e^k_{i})^{*}e^k_{j}(e^{k^{'}}_{j})^{*}e^{k^{'}}_{j^{'}}
  \Tr\left[P_{i}P_{TR}P_{j}P_{TR}P_{j^{'}}\right]\\
 & = & \sum_{ij}\left|\sum_{k}(e^k_{i})^{*}e^k_{j}\right|^{2}
 \Tr\left[P_{i}P_{TR}P_{j}P_{TR}\right] = 
 \sum_{ij}\left|\langle\vec{\epsilon}_i,\vec{\epsilon}_j\rangle\right|^2
 \Tr\left[P_{i}P_{TR}P_{j}P_{TR}\right] \,.
\end{eqnarray*}  
\end{widetext}
The trace over the projectors is easily evaluated in the basis 
$\{\ket{\varphi_i}\}$,  
$P_i=\ket{\varphi_i}\bra{\varphi_i}$, in which
$P_{TR}=\ket{\Psi}\bra{\Psi}$. It yields
$\Tr\left[P_{i}P_{TR}P_{j}P_{TR}\right] =
\left|\Braket{\varphi_i|\Psi}\right|^{2}\left|\Braket{\varphi_j|\Psi}\right|^{2}
=\left|\mu_{i}\right|^{2}\left|\mu_{j}\right|^{2}$ with
$\mu_{i}\equiv\Braket{\varphi_i|\Psi}$ and $\mu_i\neq 0$
due to total rotation, $P_iP_{TR}\neq 0 \forall i$. 
Estimating
$\left|\langle\vec{\epsilon}_i,\vec{\epsilon}_j\rangle\right|^2$ by
the Cauchy Schwartz inequality,
$\left|\langle\vec{\epsilon}_i,\vec{\epsilon}_j\rangle\right|^2 \le
\langle\vec{\epsilon}_i,\vec{\epsilon}_i\rangle
\langle\vec{\epsilon}_j,\vec{\epsilon}_j\rangle$, 
and making use of the normalization of $\vec{\epsilon}_i$,
cf. Eq.~\eqref{eq:normeps}, we obtain 
\begin{eqnarray*}
1= \Tr\left[\mathcal{D}\left(P_{TR}\right)^{2}\right] &=&
   \sum_{ij}\left|\mu_{i}\right|^{2}\left|\mu_{j}\right|^{2}
\left|\langle\vec{\epsilon}_i,\vec{\epsilon}_j\rangle\right|^2\\
&\le & \sum_{ij}\left|\mu_{i}\right|^{2}\left|\mu_{j}\right|^{2} = 1\,.
\end{eqnarray*}
In the last step, we have used 
$\sum_{i}\left|\mu_{i}\right|^{2} = 
\sum_{i}\left|\Braket{\varphi_i|\Psi}\right|^{2}
=\sum_{i}\Braket{\Psi|\varphi_i}\Braket{\varphi_i|\Psi}
= \Braket{\Psi|\Psi}=1$. Since we find one on the left hand and right
hand side, equality must hold for the inequality. Since $\mu_i\neq 0$
for all $i$, this is possible only for 
\[
\left|\langle\vec{\epsilon}_i,\vec{\epsilon}_j\rangle\right|^2  = 1\,,
\quad\mathrm{or}\,,\quad
\left|\langle\vec{\epsilon}_i,\vec{\epsilon}_j\rangle\right|  = 1\,,
\quad\forall i, j\,.
\]
Therefore, the normalized vectors $\vec{\epsilon}_i$,
$\vec{\epsilon}_j$ are identical up to a complex scalar, 
$|e^k_i|=|e^k_j|$ for all $i$, $j$ and $k$. This implies for the
proportionality constants 
$\mathcal N(k)$, Eq.~\eqref{eq:defN}, equality of all summands, 
\[
\mathcal{N}\left(k\right)=\sum_{i=1}^de^k_{i}(e^k_{i})^{*}= d \;
(e^k_{a})^{*}e^k_{a}\,. 
\]
Each component is thus given by  
$e^k_{i}=\sqrt{\mathcal{N}(k)/d}\exp\left[i\phi_{i}\right]$
which, making use of the orthogonality of the vectors $\vec{e}_k$,
$\sum_i e^k_i (e^l_i)^* \sim \delta_{kl}$, leads to  
\begin{eqnarray*}
\sum_{i=1}^de^k_{i}(e^l_{i})^{*}
&=&\sum_{i=1}^d\frac{\sqrt{\mathcal{N}\left(k\right)\mathcal{N}\left(l\right)}}{d}
\exp\left[i\phi_{i}\right]\exp\left[-i\phi_{i}\right]\\
&=&\sqrt{\mathcal{N}\left(k\right)\mathcal{N}\left(l\right)}=0
\quad \forall k\neq l\,.  
\end{eqnarray*}
For this to be true, all $\mathcal N(k)$ except one and consequently
all $E_k$ except one must be zero. By
Eq.~\eqref{eq:Ekdiag}, its representation is 
\[
E=\tilde U\left[\sum_{i} (e^1_{i})^{*}P_{i}\right]\,.
\]
Making use of $P_{i}P_{j}=\delta_{ij}P_{i}$ and $P_{i}=P_{i}^+$, 
unitarity of the time evolution follows immediately since 
\begin{eqnarray*}
E^+E & = & \sum_{i=1}^de^1_{i}(e^1_{i})^{*}P_{i}
=\sum_{i=1}^2\sqrt{\frac{\mathcal{N}\left(1\right)}{d}}P_{i}
=\sum_{i=1}^kP_{i}=\openone\,,\\
EE^+ & = &
\tilde U\left(\sum_{i=1}e^1_{i}(e^1_{i})^{*}P_{i}\right)\tilde U^+
=\tilde U\openone \tilde U^+=\openone\,,
\end{eqnarray*}
such that 
\[
\mathcal{D}\left(\rho\right)=\tilde U\rho \tilde U^+
\]
for a  unitary $\tilde U\in\text{PU\ensuremath{\left(d\right)}}$.
This concludes the proof.


\end{document}